# Evaluating QoS Parameters for IPTV Distribution in Heterogeneous Networks

Ioan Sorin COMSA*, Radu ARSINTE**

*Abstract*—The present work presents an architecture developed to evaluate the QoS parameters for the IPTV heterogeneous network. At its very basic level lie two software technologies: Video LAN and Windows Media Services with two operating systems: Windows and Linux. Three types of streams are analyzed, which will be transmitted to a Linux VLC client through means of the aggregation and access servers. The first stream is generated in real time by a capture camera, processed by the encapsulated VC-1 encoder and sent to the Media Server, while the second one is of VoD(Video on Demand) type and the third one will be handled by DVBViewer through the MPEG TS form. The first stream is transcoded in H.264-AAC such that the Linux stations will recognize its format. Through the simultaneous transmission of the three streams, we are analyzing their performance from a QoS parameters point of view by means of an application implemented in C programming language. The stream transporting the DVB-S television content was proven to ensure the best performance regarding loss of packets, delays and jitter.

*Keywords*—IPTV Content, measurements, QoS, inter-packet delay, jitter, packet loss

## I. INTRODUCTION

IPTV is defined as "multimedia services such as television/video/audio/text/graphics/data delivered over IP based networks managed to provide the required level of quality of service and experience, security, interactivity and reliability" [Wikipedia].

The IPTV service providers can offer a different number of services using their infrastructure and capacity. Some of the most common services are the real-time transmissions and video-on-demand contents (VoD) [1].

IPTV systems can offer a large number of characteristics such as: support for interactive TV (High Definition Television (HDTV), interactive games, high speed Internet navigation), time-shifting, customization (support for bi-directional communication), low bandwidth requirements, accessibility on multiple devices [1].

The high-level architecture of an IPTV environment comprises four key blocks, each one with particular functions and interdependencies [1]. The main elements of the IPTV environment are depicted in Figure 1. The first block is the content provider and involving only the information sources, the security and operating problems are not implicated. For example, the information sources may be a capture camera,

* Ing. I.S.Comsa is with Technical University of Cluj-Napoca, Romania, Communications Department, email: ionutz_comsa@yahoo.com
** Prof. R. Arsinte is with Technical University of Cluj-Napoca, Romania, Communications Department, email: Radu.Arsinte@com.utcluj.ro

local video content, or a TV channel.

The service provider block is responsible for source content, transforming it into IP content and sending it to subscribers via the network provider. The network providers assures the transport and distribution networks. They are responsible for delivering configuration, status, update and control information from the IPTV service providers to the subscribers, as well as delivering the content requested by subscribers. Subscribers (network clients) are the last element of the infrastructure, they have special equipment configured to receive, interpret and display the contents sent by the IPTV service providers, and they are bound to the license terms agreed with the IPTV service providers [1].

Once captured, the video content is converted into a digital form using the sampling, quantization and compressing processes. The most used compression methods are the type of MPEG(Moving Pictures Experts Group): MPEG-1, MPEG-2, MPEG-4, MPEG-7, MPEG-21. MPEG-2 and MPEG-4 version 10(H.264) are used by an IPTV system[2]. At the output of the encoder, the transfer rate is classified by: MPEG-CBR(Constant Bit Rate) and MPEG-VBR(Variable Bit Rate). VC-1 is a another compression technology which is adopted by the Microsoft Windows Media Video(WMV) 9 for the multimedia encoding platform [2].

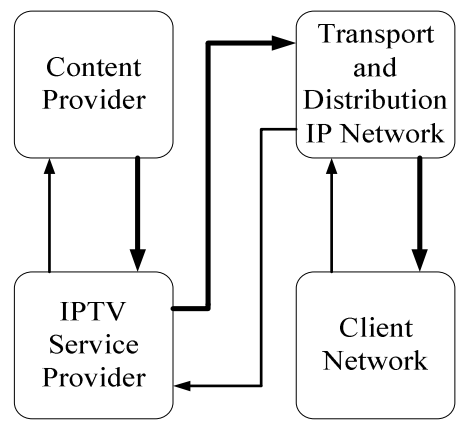

Fig. 1. IPTV high-level architecture

The packetizing and encapsulation of video content involves inserting and organizing video data into individual packets. There are a couple of different approaches to encapsulating video content, namely, MPEG over IP and VC-1 over IP. The IPTV communications model is a networking framework composed of seven (and one optional) conceptual layers that are stacked on top of each other (Figure 2) [2].

The communication process starts with the MPEG elementary stream that is outputted from the encoder. The types of information included in an elementary stream can include: frame type and rate, positioning of data blocks on screen, aspect ratio [2].

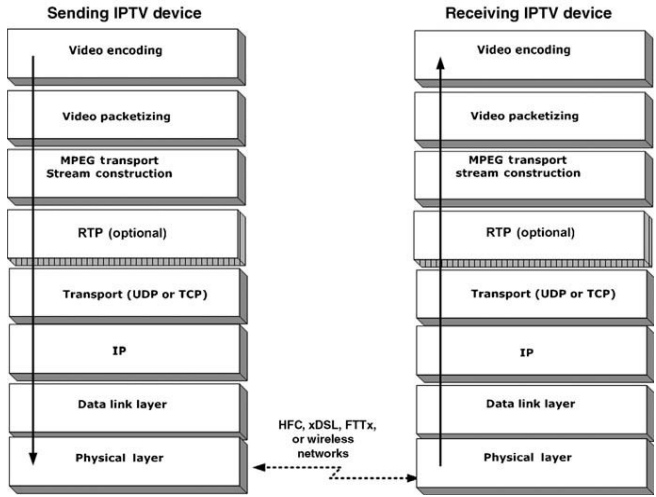

Fig. 2. The IPTV communication model [2]

In order for the audio, data, and video elementary streams to be transmitted over the digital network, each elementary stream is converted into an interleaved stream of time stamped Packetized Elementary Stream (PES) packets. A PES stream contains only one type of data from one source. A PES packet may be a fixed (or variable) sized block, with up to 65536 bytes per packet[2].

The transport stream is formed by breaking up the PES packets into fixed-sized TS packets of 188 bytes that are referenced to independent time bases. Each TS packet contains one of the three media formats, video, audio, or data.

For VC-1, the encapsulation mechanisms are similar to the MPEG. The transport mechanism involves encapsulation of VC-1 access units (AUs) inside a series of RTP packets. Each AU contains a header and variable length video payload [2].

In real-time streaming, the video content is sent using the RTP(Real-time Transmission Protocol) over UDP(User Datagram Protocol). For VoD transmission the protocol used is RTSP(Real-time Transmission Streaming Protocol) over TCP(Transport Control Protocol).

## II. QOS PARAMETERS FOR IPTV

Quality of Service represents the capacity of the network to ensure better services for a selected type of traffic. The QoS target is to assure a band allocation, to control the delay and jitter and to reduce the number of packet loss [3].

QoS parameters are divided in two parts: for real-time streaming and for VoD content. QoS parameters for real-time streaming are: traffic rate, inter-packet delay, start delay, jitter, number of lost packets, number corrupted packets, number of reordered packets.

### A. Traffic Rate

Traffic rate indicate the capacity of network to permit a type of transmission. Define the number of packets during the transmission. Dividing the number of bytes by the transmission duration we obtain the traffic rate

$$Traffic\_rate = \frac{total\_bytes}{transmission\_duration}[Bps] \quad (1)$$

### B. Inter-packet delay

Inter-packet delay represents the delay between transmitted and received packets. This parameter depends on number of nodes (routers), network traffic, routing protocols. The average value is defined in expression (2). It is important to note that all the stations must to be synchronised.

$$Interpacket\_delay = \frac{Delay_1 + Delay_2 + ... + Delay_N}{N}[s] \quad (2)$$

### C. Inter-packet jitter

Inter-packet jitter is a variation of the inter-packet delays and it is a very important parameter for real time streaming. The average jitter is defined with the formula number (3).

$$Jitter = \frac{|Jitter_1| + |Jitter_2| + ... + |Jitter_N|}{N}[s] \quad (3)$$

### D. Packet-loss parameter

The packet-loss parameter is defined in conjunction with the path between source and destination where the packet can be lost or eliminated by the router if the buffer is full [3]. If the packet is corrupted, it is declared lost. The lost of packets is depending on the current state of the network which cannot be anticipated. The algorithm used for packet-loss is the monitoring of the RTP sequence number and according to this it can make a decision if the packet between two sequence numbers is lost or not.

Some bits transmitted through the network can be corrupted. This can affect the quality if the number of corrupted bits is high. The parameter that can evaluate the number of corrupted packets is PER(Packet Error Rate) calculate with the formula (4).

$$PER = \frac{Number\_corrupted\_packets}{Number\_received\_packets} \times 100\% \quad (4)$$

The packet reordering appears when at the receiving side, the packets can arrive out-of-order because of the different paths chosen by routers. A packet is considered reordered if the sequence number is smaller than the sequence number of the previous packet received. We use the RTP sequence number of the packets for every UDP port used during transmission.

## III. THE PROPOSED ARCHITECTURE

The implementation of an IPTV network requires the use of different types of technologies. They may relate to the operating systems that are used for the stations involved. The proposed architecture (an enhanced version of the architecture proposed in [4]) is using two operating systems: Windows and Linux. In this case the IPTV network became heterogeneous.

These operating systems use the combination of another types of technologies: Windows Media Services 9 (WMS) and Video LAN(Local Area Network) Server(Client). Windows Media Services are used only by the Windows stations. VLC or VLS is running on both station types: Windows or Linux.

The main idea is based on forming an IPTV mini-network which follows the architectural model with the translation and transmission methods necessary to ensure an acceptable QoS level. Figure 3 illustrates this concept: the capture camera, the Windows Media Encoder and the Technisat DVB receiving card form the content provider block, the Windows Media Server, DVBViewer program and VLC Windows Server 1 represents the IPTV service providers, the VLC Windows server 2 and the first router take place of the transport and aggregation network and the VLC Linux server 3 with the second router can be seen as a access mini-network.

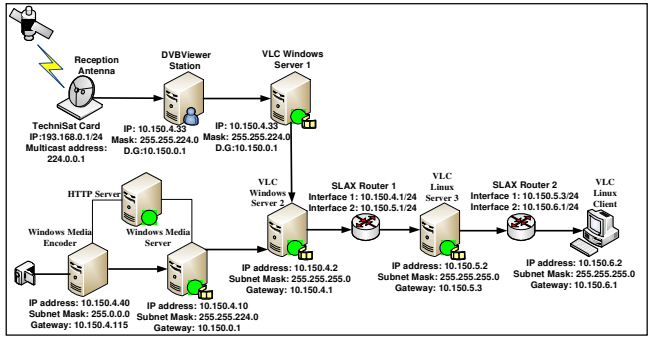

Fig. 3. The proposed IPTV test architecture

The scenario involves three types of streams. The first is taken from a capture camera, it's CBR coded by Windows Media Encoder transmitted forward to the Windows Media Server using HTTP protocol. The second stream is a VoD type and is generated by the encoder. Another stream is taken from the DVBViewer, a program that transforms the radio-frequency signals from the different types of satellites, containing a large number of channels and encode it for multicast transmission to a VLC Windows server 1.

The Windows Media Encoder with IP address 10.150.4.40/24 converts the video content into a digital form and then encapsulated it to be sent to Windows Media Server. VC-1 is the technique that is used.

Windows Media Server(IP address: *10.150.4.10/24*) is receiving the audio-video content using the HTTP and TCP protocols. The connection between encoder and server is set with the command *http://10.150.4.40/Encoder:80*. The video content is distributed by the Media Server through *Channel1*. This channel transmits in multicast type the media information to the VLC Windows Server 2. However, the VLC server 2 can access the windows media content using MMS protocol(*mms://10.150.4.10/Channel1*). This server is running at the station with *10.150.4.2/24* IP address. His role is to trans-code the WMS video content from *VC-1* format into *H.264&AAC* format, so that the output stream to be recognized on Linux stations.

The second stream is generated on the encoder station and is transmitted directly to VLC Windows Server 2 without any trans-coding techniques.

The third stream is taken from Technisat network card with *193.168.0.1/24* IP address. DVBViewer is an interface between this network card and user. The stream is transmitted from the DVBViewer station on *224.0.0.1* IP multicast address and 7792 transmission port. The interface address for DVB program is *193.168.0.1/24*.

The VLC Windows Server 1 is responsible for transforming the RTP multicast transmission into RTP unicast transmission, such as the resulting stream to be transmitted to the VLC Windows server 2 in MPEG TS format. This server is running on *10.150.4.33/24* IP address and the transmitting port is *1234*.

All of these streams are sent to the VLC Windows Server 2. The role of this server is to aggregate all the streams received and represents the first element of the private network.

All traffic is sent to the VLC Linux Server 3 through the first router. This is simulated using a station with SLAX Linux distribution. This will have two interfaces with IP address, identical with the default gateways of the VLC servers: *10.150.4.1/24, 10.150.5.1/24*. First, it adding the interface and network addresses.

    *ip addr add 10.150.4.1/24 brd + dev eth0*
    *ip addr add 10.150.5.1/24 brd + dev eth0*
    *ip route add 10.150.4.0/24 dev eth0*
    *ip route add 10.150.5.0/24 dev eth0*

The VLC Linux Server 3(*10.150.5.2/24*) takes the streams from VLC Windows Server 2. The role of this server is to forward the traffic to the VLC Linux Client through the second router with SLAX distribution. The default gateways are: *10.150.5.3/24* and *10.150.6.1/24*. For routing the packets, at the both routers, the variable */proc/sys/net/ipv4/ip_forward* is set to the value *1*.

*echo 1 > /proc/sys/net/ipv4/ip_forward*

VLC Linux Client is receiving all the traffic on 10.150.6.2/24 IP address.

On the VLC Linux Server and Client will monitor the QoS parameters with an application named *sniffer.c*. This program will measure the traffic rate, inter-packet delay, jitter and packets loss only on the last stations.

For the remaining network will analyze the traffic rate, the other parameters are negligible because of the gigabit local area network.

Sniffer.c was compiled under Linux using *gcc*. First, it initializes the parameters like filter port, interface, number of received packets, missing frames, etc. The capture interfaces are eth0 and eth1. Setting the interface, the program obtains the network and mask addresses.

After the creation of the principal thread, the types of information printed in three files are: the index of received packet and the time stamp, the number of lost packets at every time stamp and the accumulated values of lost packets at the moment when the lost is detected.

The packet loss is detected using the RTP sequence number. The difference between two RTP sequences is printed in the packet loss file. If the difference is not 1, then the loss of packets is detected and the values are printed at the moment of time when the last RTP sequence is captured by the interface.

The measurement of the one way delay parameter implies very good clock synchronization between the server and the receiver, because the delay is obtained by comparing the time stamps of the sent and received packet. Any synchronization problem between the two elements leads to erroneous values. The VLC Linux server 3 and client are synchronized using the ntp.conf file. With it, the client takes the time in h/m/s/ms format from the server. By making the difference between the time stamps when the packets are sent by the server and the time stamps when the packets are received on the client side, we obtain the inter-packet delay and the jitter.

## IV. EXPERIMENTAL RESULTS

Experimental results are obtained by the simultaneous transmission of the streams described in the previous paragraph. All the streams are started by the generating stations and programs and are sent to the VLC Windows Server 2. This server transmits the streams to the VLC Linux Sever 3 on the following ports: the stream that is generated by the capture camera is sent on the port 5000, the second one on the port 5001 and the stream generated by DVBViewer is transmitted on the port 5002. At the VLC Linux client the streams are received on 1240, 1241 and 1242. So, the pairs of ports where are evaluated the QoS parameters are: 5000-1240, 5001-1241 and 5002-1242 corresponding to the streams in the order mentioned before.

On the first stream, the parameters for Media Encoders are:
- Audio-video encoding mode: CBR (Constant Bit Rate)
- Buffer length: 1 second;
- Video smoothness: 70;
- Frame rate: 29,97 fr/s;
- Video size: same to the input video;
- Video rate:1200kbps;
- Video codec: Windows Media Video 9;
- Audio format: 64kbps, 48kHz, stereo CBR;

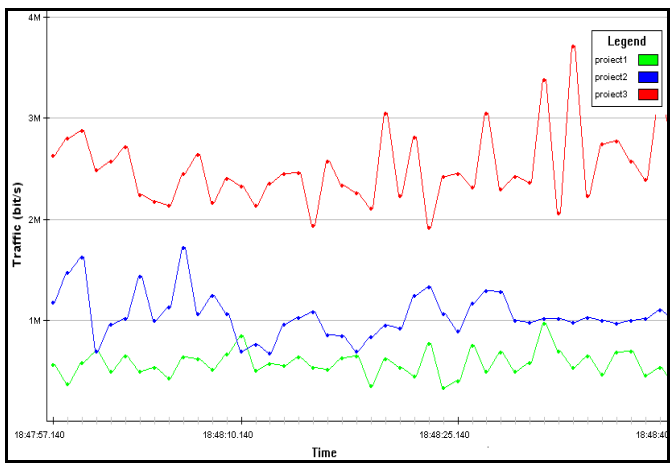

Fig 4. Traffic rate on the VLC Windows Server 2

- Audio codec: Windows Media Audio 9;
- Total rate: 1273,03kbps
- Encoded packet size: 1400 bytes;

On the VLC Windows Server 2 the traffic rate for these streams are indicated in Figure 4. The average traffic rate for the stream generated by the capture camera (green line) is 0,7Mbps. For the second stream, generated locally by the encoder, the average value is 0,9Mbps (blue line) and for the stream generated by the DVBViewer the average traffic rate is 3,7Mbps (red line).

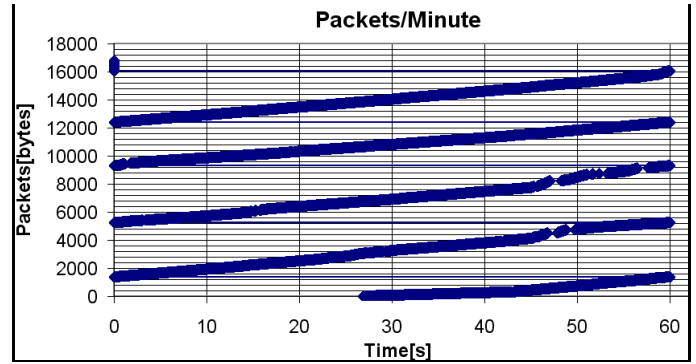

Fig 5. Traffic rate on 5000 and 1240 ports

First we evaluate the transmission rate between the VLC Linux server 3 and client. For the first stream the number of transmitted packets is 16000. Within one minute the number of packets that are transmitted from 5000 to 1240 ports is 4000 (Figure 5). Traffic rate can be calculated with formula (5). A packet contains 1372 bytes.

$$D = \frac{Number\_of\_packets}{Minute} = \frac{4000 \times 1372 \times 8}{60} = 0,7 Mbps \quad (5)$$

For the second stream, the number of transmitted packets from the port 5001 to the port 1241 in 5 minutes is 29000. Within one minute the number of transmitted packets is 5000. The rate is determined with the formula (6).

$$D = \frac{Number\_of\_packets}{Minute} = \frac{5000 \times 1370 \times 8}{60} = 0,9 Mbps \quad (6)$$

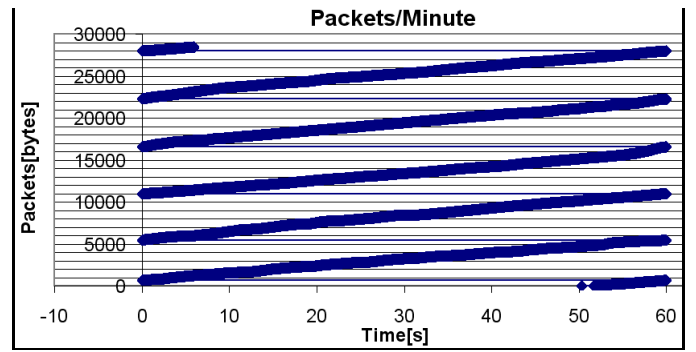

Fig. 6. Traffic rate for the second stream

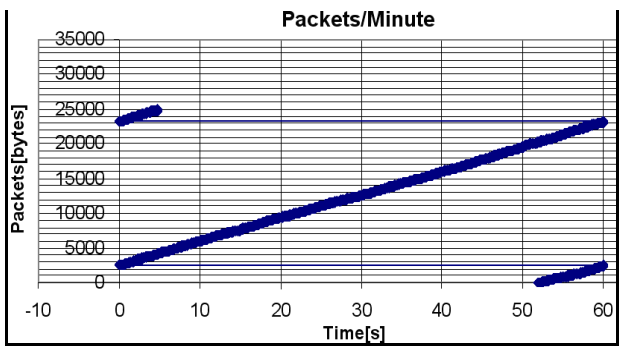

Fig. 7. Traffic rate for the DVBViewer stream

The number of transmitted packets for the third stream in one minute is 20000. Furthermore, the computed rate is 3,7 Mbps (equation 7).

$$D = \frac{20000 \times 1372 \times 8}{60} = 3,7 Mbps \quad (7)$$

Figures 6 and 7 are presenting the same parameters (traffic rate) for the second and third streams.

To determine the inter-packet delay and jitter we are considering the first 10000 transmitted packets for all the streams.

For the associate transmission of 5000 and 1240 ports, the inter-packet average value is 2,83204s, the maximum value is 5,9112452s and minimum is 0,2219906s (figure 8).

The average value obtained for jitter is 0,00511228, the minimum is -4,0012545s and the maximum 4,0012545s (figure 9). The negative value is caused by the difference between the minimum and maximum value of the inter-packet delay.

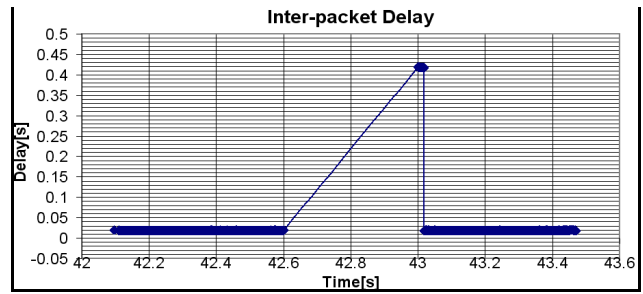

Fig. 10 Inter-packet delay for the second stream

For the local generated stream the maximum value for inter-packet delay is 4,1904033s, the average is 0,2711245s and the minimum delay is 0,1744169s. The average jitter is 0,000098s, the minimum and maximum values are: -3,9998893s and 4,0008256s (figures 10 and 11). The minimum value of the delay is smaller than the value obtained at the first stream. The same it happens with the maximum value.

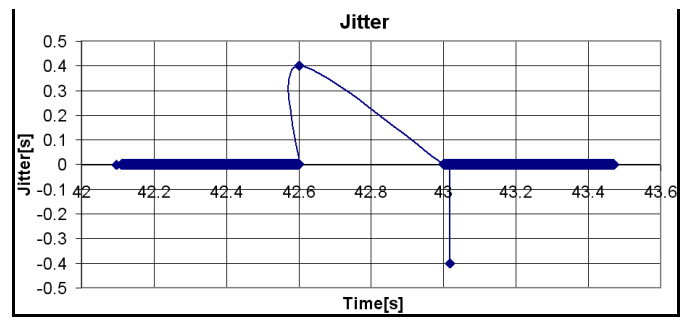

Fig. 11 Jitter for the second stream

At the transmission of the stream generated by the application DVBViewer an inter-packet delay average value of 0,2156974s is obtained, and the maximum and minimum values are 0,2541312 and 0,1851437, respectively. At the jitter the average, minimum and maximum values are: 2,75858E-5s, -0,2343453s and 0,0179553s, respectively (figure 12 and 13).

This type of stream offers a better performance than the second stream from inter-packet delay and jitter points of

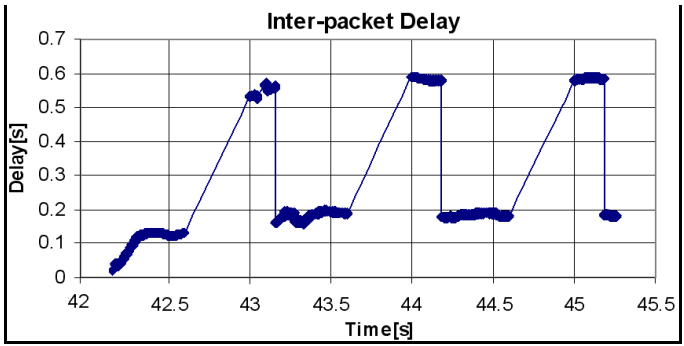

Fig. 8 Inter-packet delay for the first stream

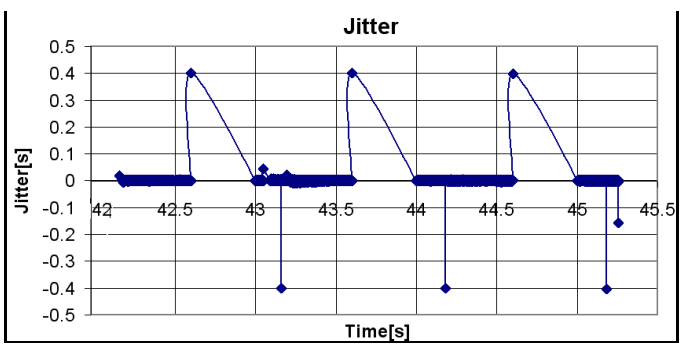

Fig. 9 Jitter for the first stream

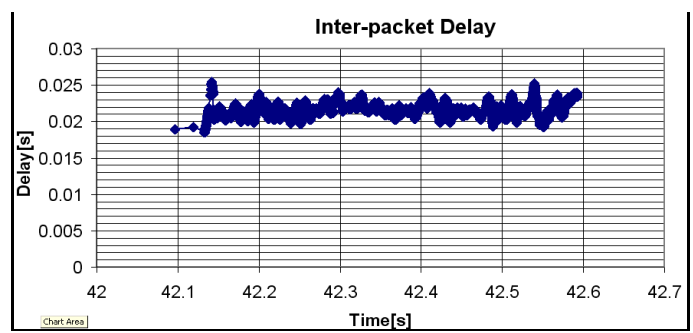

Fig. 12 Inter-packet delay for the third stream

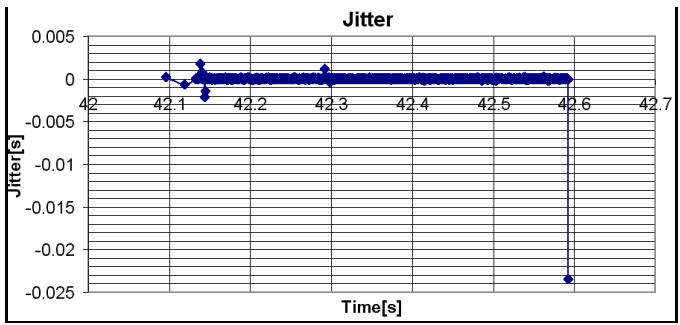

Fig 13 Jitter for the third stream

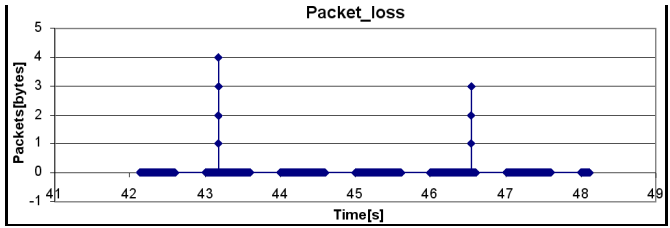

Fig. 14 The number of lost packets on port 5000

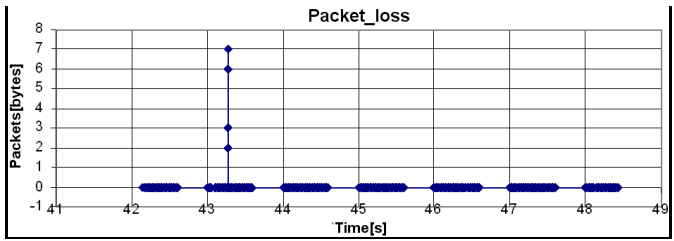

Fig.15 The number of lost packets on port 1240

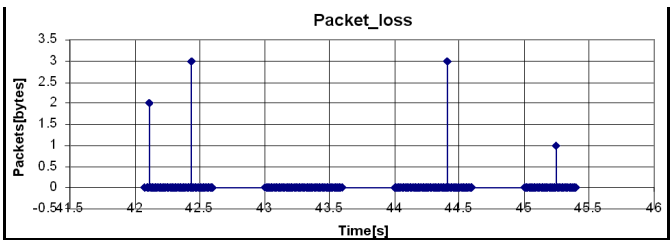

Fig. 16 The number of lost packets on port 5001

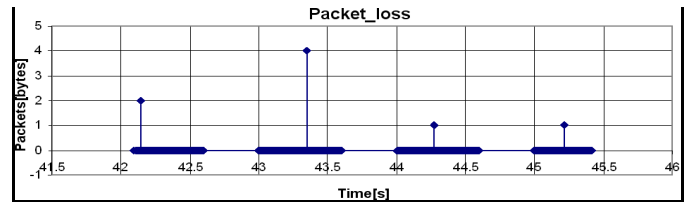

Fig. 17 The number of lost packets on port 1241

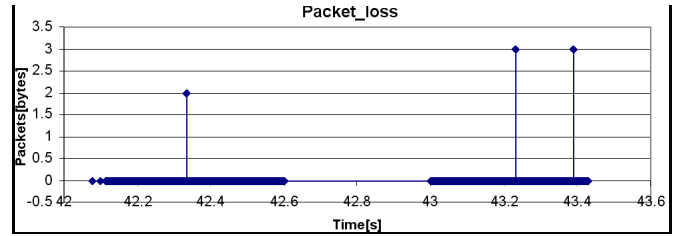

Fig.18 The number of lost packets on port 5002

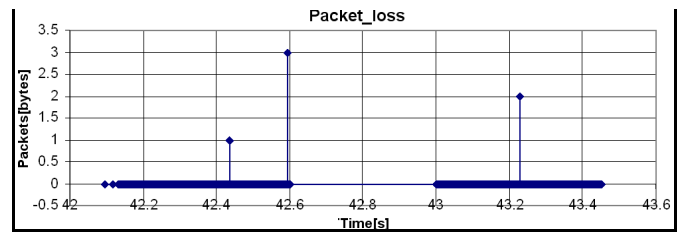

Fig. 19 The number of lost packets on port 1242

view.

To determine the number of lost packets, we consider the number of received packets for each port of the respective stream. In this test scenario, the duration of the transmission for all streams is 8 minutes.

The received packets on port 5000 are 26436 corresponding to a duration of 8 minutes. The total number of lost packets is 100, representing a percentage of 0,38%. For a sample of 20000 packets, the number of lost packets is 16 (figure 14).

On port 1240 at the VLC Linux client, the number of the received packets is 24864 of which 74 are declared lost (percentage of 0,3% from total received packets). On a sample of 20000 received packets the number of lost packets is 18 (Figure 15).

Within 8 minutes, on port 5001, the number of received packets is 44 386 and 71 are declared lost. It represent 0,16% percentage of a total received packets. On the sample of 20000 received packets, 10 packets are lost (Figure 16).

The number of received packets on port 1241 is 44361 and 63 are lost representing a 0,14% percent from the total number of received packets. For a sample of 20000 received packets, 8 are lost (figure 17).

The port 5002 receives 106 039 packets and 57 are declared lost (a percentage of 0,03%). Figure 18 illustrates the number of lost packets on a sample of 20000 received packets. Figure 19 is presenting the same parameter for port 1242.

## V. CONCLUSIONS AND FURTHER WORK

For a final overview of the results, in Table 1 we are marking the worst case (blue) and the best performance (red) of all streams from QoS parameters point of view. White cells of the table represent neutral values.

The first stream, generated by the camera stream capture, offers the worst results for all QoS parameters. The traffic rate is 0,7Mbps. For this stream, part of information is lost because of H.264 transcoding process performed in VLC Windows Server 2.

For the second stream generated locally on the encoder station, is obtained the minimum of inter-packet delay (0,1744169s).

The third stream generated by DVBViewer offers the best values for maximum and average inter-packet delay, maximum, minimum and average jitter and the percentages of lost packets on ports 5002 and 1242 are the smallest. This can be easily explained, since digital video broadcasting content is already MPEG encoded and packetized.

TABLE I
STREAMING PERFORMANCE FROM QoS POINT OF VIEW

| Streams/QoS Parameters | Stream 1 | Stream 2 | Stream 3 |
|---|---|---|---|
| Rate | ■ | - | ■ |
| Maximum interpacket delay | ■ | - | ■ |
| Average interpacket delay | ■ | - | ■ |
| Minimum interpacket delay | | ■ | - |
| Maximum interpacket jitter | ■ | - | ■ |
| Average interpacket jitter | ■ | - | ■ |
| Minimum interpacket jitter | ■ | - | ■ |
| Percentage of lost packets on port 5000 | ■ | - | - |
| Percentage of lost packets on port 5001 | - | - | - |
| Percentage of lost packets on port 5002 | - | - | ■ |
| Percentage of lost packets on port 1240 | ■ | - | - |
| Percentage of lost packets on port 1241 | - | - | - |
| Percentage of lost packets on port 1242 | - | - | ■ |

Similar systems and comparable results are presented in [8], and this is encouraging us to continue the development of the evaluation system.

This study is showing clearly that the power of general purpose computers is limited, when complex manipulations of information, like transcoding, are needed. In this case dedicated hardware (based on ASICs or FPGAs) must be employed, to ensure a proper quality for IPTV services.

Future work will be dedicated to more test scenarios, implemented in an even more real IPTV environment, with multiple users and different streaming applications. It is possible to add the evaluation not only of the streaming parameters, but also for additional elements, specific to IPTV applications. Such elements are suggested in work [9].

A real challenge will be to evaluate the IPTV QoS in physical networks different from Ethernet, like WiFi (suggested in [10]) or ADSL [11]. Reference [12] contains several test scenarios and a large number of measurements.